\documentclass[amsmath,amssymb,aps,prb,
				twocolumn,groupedaddress,
                showkeys,letterpaper,10pt]{revtex4-2}

\usepackage{orcidlink}
\usepackage{datetime}
\usepackage{graphicx}

\graphicspath{{./}{./figs/}}

\begin{document}

\date{\today}

\newlength{\figwidth}
\setlength{\figwidth}{0.95\linewidth}

\title{Thermal conductivity of 3C-SiC from configuration space sampling}

\author{Paweł T. Jochym\,\orcidlink{0000-0003-0427-7333}}
\email{pawel.jochym@ifj.edu.pl}
\affiliation{Institute of Nuclear Physics,
Polish Academy of Sciences, Cracow, Poland}

\author{Jan Łażewski\,\orcidlink{0000-0002-7585-8875}}
\affiliation{Institute of Nuclear Physics,
Polish Academy of Sciences, Cracow, Poland}

\begin{abstract}
Cubic silicon carbide phonon thermal conductivity has been calculated using anharmonic
phonon analysis. The atomic interaction model was built using displacement-force data
obtained with the High Efficiency Configuration Space Sampling (HECSS) technique and
density functional theory calculated forces. In the new version of HECSS, we replaced
the Markov chain scheme of Metropolis-Hastings Monte-Carlo with weighting of the final
sampling according to the target distribution. This increased the efficiency of the
method and allowed us to use all generated samples. 
The quality of the proposed method is confirmed by the accuracy with which 
the experimental results taken from the literature were reproduced.
\end{abstract}

\maketitle

\section{Introduction}
\label{sec:intro}
Silicon Carbide (SiC) can crystallize in a rich family of
polytypes. It is quite exceptional, having more than 250 possible
structures identified \cite{cheung_silicon_2006,
kelly_correlation_2005, daulton_polytype_2002}. The stoichiometry of
all polytypes is the same (50:50 proportion of Si and C); the only
difference is the stacking order and arrangement of consecutive atomic
layers. This results in cubic, hexagonal, or rhombohedral structures
as well as lower symmetry polytypes with very large lattice vectors
(even above 300~nm \cite{kelly_correlation_2005}). Despite this large
spectrum of structures, only hexagonal (4H-SiC, 6H-SiC) and cubic
(3C-SiC) are commercially available. The structure determines the
properties of the material, but since the composition of all
polytypes is the same, they all fall in the same ballpark. All
structures are classified as wide band gap semiconductors, with a gap
energy of 2.3-3.2~eV, which is much higher than silicon. At the same
time, they have a large breakdown field (200-400~MV/m). These
properties result in low on-state resistance and, simultaneously, low
leakage current. Combined with high saturated electron velocity, which
is important for high-frequency behavior, and thermal conductivity
approximately three times higher than silicon, all these
properties make silicon carbide a promising material for all kinds of
industrial applications, particularly for devices working in
high-temperature environments, and a potential alternative to
currently used semiconductors (e.g. gallium nitride). What is more,
SiC is also known to be biocompatible \cite{saddow_silicon_2011,
saddow_3csic_2014} and radiation resistant \cite{nava_radiation_2003}.

The material is not without some drawbacks, the major one being the
difficulty of growing large-volumes good-quality single crystals,
similarly to silicon monocrystals. Although 3C-SiC epilayers can be
grown on silicon wafers
since the early 1980s \cite{severino_highquality_2010,
nishino_production_1983, nishino_epitaxial_1987,
liaw_epitaxial_1985, steckl_epitaxial_1992, nordell_design_1996,
kordina_growth_1995}, they are still of relatively low
crystallographic quality, which limits their applications and favors
more expensive hexagonal bulk SiC wafers in the fabrication of devices
requiring larger volume and/or higher crystal quality.

The electrical properties mentioned above make silicon carbide a
promising material for the production of high-power,
high-temperature devices. The application range is, however, not
limited to these fields. Silicon carbide's electro-optical
properties, such as its large second-order ($\chi^2$) susceptibility
\cite{wu_secondharmonic_2008} and high third-order non-linearity
($\chi^3$) \cite{cardenas_optical_2015,martini_four_2018}, make
it also valuable for optical frequency conversion devices
\cite{lukin_4hsiliconcarbideoninsulator_2020}.
This is particularly because of its
aforementioned favorable properties, such as high
chemical resistance \cite{daves_amorphous_2011}, hardness
($E\approx 450$~GPa), and high thermal conductivity
\cite{jackson_mechanical_2005}. This last property plays a
significant role in high-power applications since it helps to quickly
remove the dissipated heat from the device using the substrate as a
heat conductor instead of elaborated heat transport solutions 
(e.g.~bonded copper heat spreaders which need to be electrically insulated
from the device, while being in good thermal contact with the
heat-producing active parts).

The ability to accurately predict thermal conductivity for a novel
material or structure using \textit{ab initio} computation has been
one of the goals for computational material science. While the
theoretical framework of heat transport in crystalline solids is well
understood \cite{peierls_zur_1929,simoncelli_unified_2019}, its
practical application encounters numerous difficulties. Firstly, heat
transport is connected with phonon interactions and energy exchange
between vibration modes. Thus, it is an intrinsically \textit{anharmonic} process
because harmonic phonons do not interact. Secondly, the interaction
model used to construct lattice dynamics should correspond to a
physically realistic state of the system, which is a thermodynamic
equilibrium at a given temperature. Thus, we need to build an
anharmonic interaction model for the system. Such a model requires a
large set of parameters -- interatomic force constants (IFCs), which
are often obtained from molecular dynamics
\cite{hellman_lattice_2011, hellman_temperature_2013,
hellman_temperaturedependent_2013}. This approach, while quite
expensive computationally, simultaneously solves the second of the
above difficulties by deriving data for the interaction model from the
representation of the thermal equilibrium provided by the molecular
dynamics. In our work, we have used a similar approach of
constructing the interaction model from the representation of the
thermal equilibrium state, but we have replaced the molecular dynamics
with the recently introduced HECSS method \cite{jochym_high_2021}
which samples the thermodynamic ensemble with high efficiency. The
anharmonic lattice dynamics model is then used to calculate the
temperature-dependent heat conduction coefficient for the finite 3C-SiC
crystal.

\section{Calculation methods}

The determination of the thermal conductivity of the crystal
necessitates the development of an anharmonic lattice dynamics model
that corresponds to the thermal equilibrium. Traditionally, this is
done with the displacement-force data set generated with
\textit{ab initio} molecular dynamics \cite{hellman_temperature_2013}.
This method is computationally expensive and quite wasteful
-- since one throws away most of the data computed along
the trajectory. Instead, we have used the thermodynamic ensemble
sampling technique (HECSS) based on the generalized equipartition
theorem. The details of the approach are described in our earlier
work\cite{jochym_high_2021}. This approach was already successfully
used to investigate negative thermal expansion
\cite{jochym_influence_2022}, structural phase transitions
\cite{ptok_dynamical_2022, pastukh_anharmonicity_2023}, orbital order
\cite{ptok_electronic_2021} and chiral phonons
\cite{ptok_chiral_2021}. These results demonstrate that it is
possible to effectively generate samples providing an appropriate
representation of the system in thermal equilibrium at a given
temperature and build an adequate lattice dynamics model for the
material extending to the analysis of negative thermal expansion
effects\cite{jochym_influence_2022}.

Here, we present a series of calculations for 3C-SiC using
VASP as a source of energies and forces, using a range of system
sizes (from $2^3$ to $5^3$ supercell) and a standard PAW-PBE calculation
setup\cite{kresse_initio_1994, kresse_efficiency_1996,
kresse_efficient_1996, perdew_generalized_1996}. The aim of our work
is to calculate phononic thermal conductivity based on anharmonic
lattice dynamics derived from the DFT data. This goal requires
substantially more data points, compared to
standard lattice dynamics calculations, due to the large number
of independent interatomic force constants (IFCs) of the third and
fourth order required to derive an anharmonic lattice dynamics model.

The DFT calculations used standard PAW-PBE atomic datasets with energy
cutoffs increased by 30\% above the standard value
(i.e., 'Accurate' mode of VASP), $\Gamma$-centered k-space sampling
grid with a maximum spacing of 0.25~{\AA} and $\Delta E=10^{-7}$~eV
iteration stopping criterion. The HECSS procedure was used to generate
enough data points to obtain 5\% convergence of the phonon frequencies
derived from the fitted IFCs. The IFC fitting procedure employed
ALAMODE\cite{tadano_2014,tadano_2015,tadano_2018,oba_2019} code to
fit the interaction model and calculate lattice dynamics parameters
(phonon frequencies and lifetimes) of the investigated crystal.

\subsection{Distribution shaping}

\begin{figure}
\centering
\includegraphics[width=\figwidth]{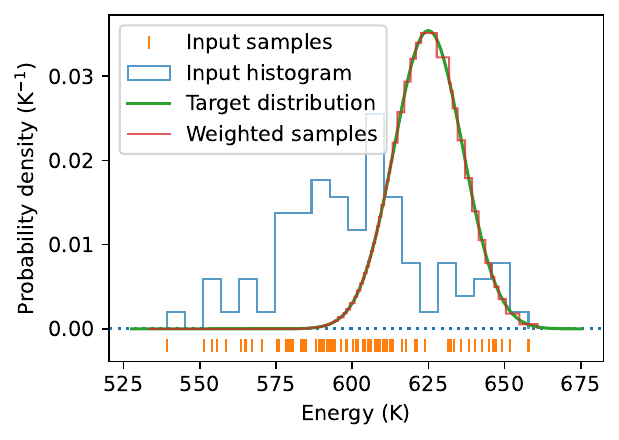}
\caption[]{DFT-calculated potential energy distribution generated with
HECSS for target temperature $T=600$~K reshaped to artificial
distribution (green line), centered at 625~K and squeezed by the
factor of 2. The resulting step-wise approximation of the target
distribution is plotted as a red line. Orange lines at the bottom
indicate the energies of the input samples.}
\label{fig:reshape}
\end{figure}

Faithful representation of a thermodynamic ensemble for the system in
thermal equilibrium requires generating an accurate
sampling of the configurations of the system with a correct, Gaussian,
energy distribution centered around $\bar{E}=3 k_\mathrm{B} T/2$ (in
three dimensions)\cite{jochym_high_2021}. This is almost
automatically achieved for large systems, thanks to the central limit
theorem, but for typical systems used in DFT calculations of tens of
atoms must be carefully observed and verified.

The standard HECSS procedure, described in our earlier
work\cite{jochym_high_2021}, generates the energy probability
distribution using the standard Metropolis-Hastings algorithm. This
algorithm is an established cornerstone of the probability
distribution sampling field. However, it requires a large number of
samples to achieve high-quality sampling of the distribution. To
improve the efficiency of the whole procedure, we have developed a
simple weighting procedure based on a few observations applicable to
energy distribution for systems in thermal equilibrium (but not
necessarily to other distributions):

\begin{itemize}
\item \emph{Any} atomic configuration is possible; thus, all generated
configurations can be included in the sampling.
\item The proper representation of thermal equilibrium requires the
assignment of appropriate probabilities (weights) to the
configurations.
\item The Metropolis-Hastings generates the weights by multiplication
of samples with the largest probabilities.
\item The weights may be calculated directly, provided we know the
full formula for the target distribution.
\item In the case of thermal equilibrium, we know all required
parameters (i.e., functional shape, its parameters, and norm).
\end{itemize}

The resulting sampling may be understood as a step-wise approximation
of the target distribution, as illustrated in Figure~\ref{fig:reshape}
using actual DFT data from the system of 512 atoms (4$\times$4$\times$4
supercell) at 600~K and an artificial target distribution which
is, for illustration purposes, shifted by 25~K and squeezed by a
factor of 2 in energy.

The result of this procedure is a substantial increase in the
effectiveness of the distribution sampling and improved elasticity of
the whole procedure. The main benefits are:
\begin{itemize}
\item All data points are used in the final sampling with weights
adjusted according to the target distribution.
\item It is possible to change the parameters of the final
distribution without recalculating of any input (DFT) data points.
\item When the input data set spans a range of energies, it is
possible to generate samplings for any temperature corresponding to the
whole energy range (i.e., scan the temperature range) without any
recalculation of DFT input data.
\end{itemize}

\subsection{Lattice dynamics}

Since the described method is a modification of the original HECSS
procedure \cite{jochym_high_2021}, we need to check if the convergence
properties still hold true. The final distribution is properly
generated in the new procedure by construction. Thus, we have decided
to use lattice dynamics, derived from the model constructed based
on the generated sampling, as a validity check for the new procedure.
We have selected the RMS difference between frequencies
evaluated on the 20$\times$20$\times$20 grid with models derived from
the varying number of samples --- randomly selected from the whole
sampling --- against the frequencies calculated from the full set of
samples.

\begin{figure}[!htbp]
\centering
\includegraphics[width=\figwidth]{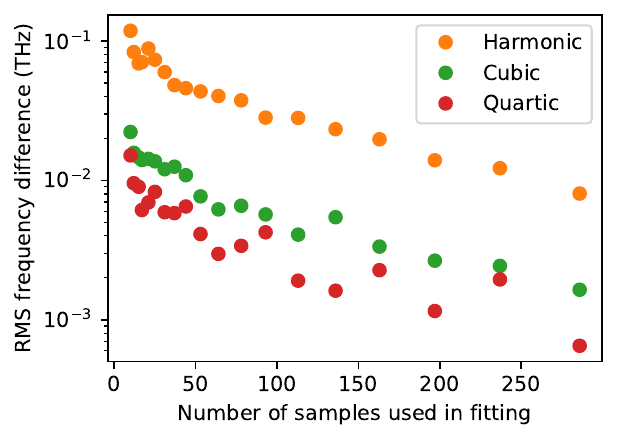}
\caption[]{Convergence of phonon frequencies with respect to the number
of samples. The samples were randomly selected from the full set of
weighted samples from the full HECSS procedure.}
\label{fig:frq_conv}
\end{figure}

The results of this comparison executed for harmonic, cubic, and
quartic models are presented in Figure~\ref{fig:frq_conv}. The
difference between harmonic and non-harmonic models is clearly visible
there. The harmonic model, which is the simplest approximation, shows
the strongest dependence on the number of samples, while anharmonic
models --- probably better fitting the interaction potential
--- show much smaller variability. This effect appears despite
a much smaller number of parameters of the harmonic model 
compared with cubic or quartic models.

\begin{figure}[!htbp]
\centering
\includegraphics[width=\figwidth]{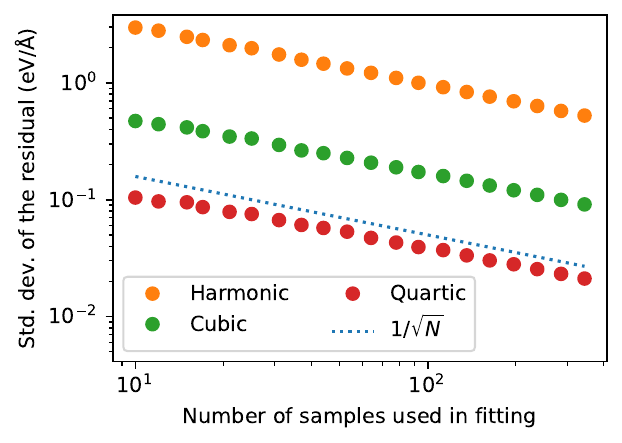}
\caption[]{Standard deviation of the residual of the models with
respect to number of used samples. The samples were randomly selected
from the full set of weighted samples from the full HECSS procedure.
The dotted line represents a function proportional to $1/\sqrt{N}$,
where $N$ is the number of samples used in the fit. }
\label{fig:rms_res}
\end{figure}

The comparison of RMS residuals from the models presented in
Figure~\ref{fig:rms_res} further confirms the above conclusion and
directly shows significantly smaller residuals of the higher-order
models, i.e., a much better representation of the input data by the
anharmonic models (cubic and quartic). The linear, in the log-log
graph, character of the RMS curves in Figure~\ref{fig:rms_res} should
be attributed to statistical reasons. It is only expected that the
standard deviation of the random variable falls inversely
proportionally to the square root of the number of samples $\sqrt{N}$
(dotted line in Figure~\ref{fig:rms_res}). Thus, this aspect of the
data is not specific to the presented case. On the other hand, the
relative quality of the models is represented by the relationship
between their trend lines, showing well over one order of
magnitude difference between harmonic and quartic models and
indicating non-negligible anharmonic component in the data.

The standard procedure for extraction of the interatomic force
constants (IFC) involves fitting the parameters of the
model potential to the displacement-force data provided by the HECSS
procedure described above. The shape of the model potential is defined
by its order - in this work we have used a fourth-order potential
which corresponds to four-body interactions in the system. The form of
the potential is further determined by the range of the interactions
(two-, three-, and four-body) specified by the cutoff distances,
separately for each order of interaction and type of interaction
(e.g., C-C, C-Si, etc.). The implementation of the TDEP scheme
(ALAMODE\cite{tadano_2014,tadano_2015,tadano_2018,oba_2019}) used in
this work provides great flexibility in this regard. The quality of
the fit strongly depends on the selection of the order and cutoff
radii for the interaction model. Unfortunately, the computational
cost of the fitting procedure grows very fast with the expansion of
the cutoffs, due to the very fast growth of the parameter space of the
potential. For the 8 and 6~{\AA} cutoffs, for third- and
fourth-order IFCs respectively, the number of free parameters exceeds
30,000.

\begin{figure}[!htbp]
\centering
\includegraphics[width=\figwidth]{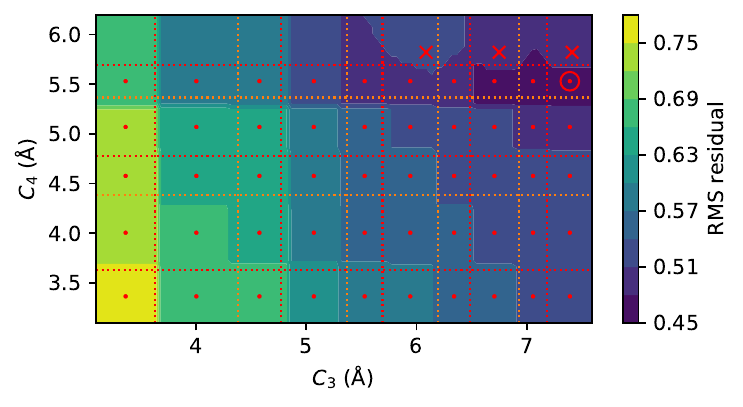}
\caption[]{Relative RMS fitting errors as a function of three- and
four-body cutoffs used in the fitting procedure. Red dots mark a
combination of cutoffs used in test runs, the color map is a nearest
neighbor interpolation of data obtained at those points. Red x's mark
calculations with expanded C-Si interaction range. The selected
combination of cutoffs is marked with an open red circle. Red and
orange dotted lines mark distances of consecutive coordination shells
in the structure: C-Si (red) and C-C, Si-Si (orange). See also
description in the text.}
\label{fig:cutoff}
\end{figure}

To select an optimal set of cutoffs providing high accuracy, we have
executed a series of test calculations on 900~K dataset with
various parameters. A summary of these calculations is presented in
Figure~\ref{fig:cutoff} as a map of the fit residuals produced by the
third- and fourth-order cutoffs ($C_3$ and $C_4$ respectively).
Considering the important role anharmonic interactions play in thermal
conductivity phenomena, we have selected cutoffs of 7.5 and
5.5~{\AA} for third and fourth-order IFCs, respectively. The selected
parameters are marked with an open circle in Figure~\ref{fig:cutoff}.
The selection was guided by the quality of the obtained models and the
positions of the coordination spheres in the system (red/orange dotted
lines in Figure~\ref{fig:cutoff}) and led to 19~000 IFCs in
the used interaction model. The quality of the constructed model is
indicated by the value of residuals, which are 0.46 for the full
anharmonic model, relative to 11.36 for the harmonic model.

\begin{figure}[!htbp]
\centering
\includegraphics[width=\figwidth]{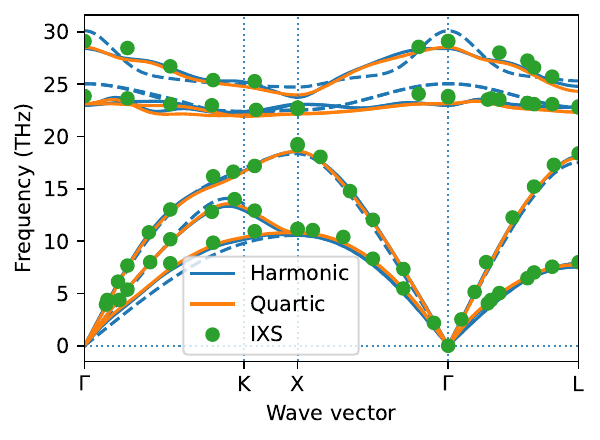}
\caption[]{Phonon dispersion relation calculated with harmonic (solid
and dashed blue line) and quartic models (orange solid line) compared
to experimental data from inelastic X-ray
scattering\cite{serrano_determination_2002}. The dashed blue line
shows the harmonic model derived using interaction cutoff
$C_2=3.7$~{\AA}.}
\label{fig:phonons}
\end{figure}

The vibrational modes in the system depicted in
Figure~\ref{fig:phonons} were calculated from the IFCs of the model
using standard dynamical matrix diagonalization implemented in the {\sc
alamode} package. The Born effective charges, required for
non-analytical term correction, were taken from the measurements of
Olego, Cardona, and Vogl\cite{olego_pressure_1982b}.

The contrast between the quality of harmonic and anharmonic models
suggests the important role of anharmonicity in the description of the
system. Furthermore, substantial change in the fits' residuals along
the $C_4$ axis in Figure~\ref{fig:cutoff} indicates a
non-negligible role of fourth-order anharmonicity. However, as is
clearly visible in Figure~\ref{fig:phonons}, its influence on
lattice vibrations is small and limited mostly to optical branches
near the X point in the Brillouin zone.

The comparison with experimental data obtained with inelastic x-ray
scattering\cite{serrano_determination_2002} and included in
Figure~\ref{fig:phonons} (green dots) demonstrates that even the
harmonic model (blue line) fits well with experimental frequencies if
the interaction range is large enough. The effects of shortening of
the interaction range are demonstrated by the dashed blue line in
Figure~\ref{fig:phonons} calculated with a harmonic interaction
cutoff reduced to $C_2=3.7$~{\AA}.

Small differences in frequency terms between harmonic and quartic
models visible in Figure~\ref{fig:phonons} indicate a small influence
of fourth-order anharmonicity on the vibrational frequencies in this
material. To verify this conclusion, we performed a series of
self-consistent phonon (SCPH) calculations using the final, best-fit
model for the temperatures between 200~K and 1200~K. The maximum RMS
difference between renormalized SCPH frequencies at extreme
temperatures turned out to be consistent with zero ($\approx 0.1 \pm
0.1$~THz), validating the above conclusion. Thus, the only remaining
significant anharmonic part of the model is the cubic term, which
mainly influences the lifetimes of the phonons in the system.

\subsection{Phonon lifetime and lattice thermal conductivity}

The goal of this work is the calculation of the
lattice contribution to the thermal conductivity of the 3C-SiC
crystal. We have estimated the lattice thermal conductivity tensor
using the relaxation time
approximation\cite{peierls_zur_1929,sun_lattice_2010,tadano_2018}, as:
\begin{equation}
\kappa_{\mathrm{ph}}^{\mu\nu}(T) = \frac{1}{V N}
\sum_{\boldsymbol{q},j}c_{\boldsymbol{q}j}(T)v_{\boldsymbol{q}j}^{\mu}v_{\boldsymbol{q}j}^{\nu}\tau_{\boldsymbol{q}j}(T),
\label{eq:kappa}
\end{equation}
where $\tau_{\boldsymbol{q}j}(T)$ is a $j$-th branch phonon lifetime
at point $\boldsymbol{q}$ in the reciprocal space and temperature
$T$. Mode group velocities $\boldsymbol{v}_{\boldsymbol{q}j} =
\frac{\partial \omega_{\boldsymbol{q}j}}{\partial \boldsymbol{q}}$ are
evaluated in {\sc alamode} by the central difference formula. Finally,
$V$ is the unit cell volume, $N$ is the number of unit cells in the
crystal, and mode heat capacity is $c_{\boldsymbol{q}j} =
\hbar\omega_{\boldsymbol{q}j}\partial n_{\boldsymbol{q}j}/\partial
T$, where mode energy is proportional to its frequency
$\omega_{\boldsymbol{q}j}$ and Bose-Einstein distribution is
$n_{\boldsymbol{q}j} = 1/(e^{\hbar\omega_{\boldsymbol{q}j}/kT}-1)$.

Phonon lifetime $\tau_{\boldsymbol{q}j}(T)$ is derived from mode
linewidths connected with anharmonic interactions
$\Gamma_{\boldsymbol{q}j}^{\mathrm{anh}}$ and isotope scattering
$\Gamma_{\boldsymbol{q}j}^{\mathrm{iso}}$ with Matthiessen’s formula:
\begin{equation}
\tau_{\boldsymbol{q}j}^{-1}(T) = 2
\left(\Gamma_{\boldsymbol{q}j}^{\mathrm{anh}}(T) +
\Gamma_{\boldsymbol{q}j}^{\mathrm{iso}}\right).
\end{equation}

The components of the phonon linewidth in the above equation are
anharmonic processes, mainly third-order components of the interaction
model\cite{tadano_2014,tadano_2018,oba_2019} and isotope scattering
included following the mass perturbation
approach\cite{tamura_1983}.

The results presented above indicate a rather quick convergence of the
harmonic model with the size of the sample and interaction cutoff. This
is not very surprising, due to the close to linear relationship
between displacements and forces acting on atoms in the 3C-SiC crystal.
In contrast, a higher-order model cannot benefit from such
circumstances, since it requires a large number of parameters (orders
of magnitude higher than the harmonic case) which are determined by
higher derivatives of the energy surface. Thus, it is only expected
that quantities such as phonon lifetimes or the thermal conductivity
of the material may be much more sensitive to subtle features of the
sampling used and require larger samples to
converge\cite{jochym_high_2021}.

\section{Results}

As described above, the thermal conductivity is necessarily connected
with phonon-phonon and phonon-lattice interaction. The first
phenomenon facilitates the distribution of energy over all degrees of
freedom, and the second mainly dissipates phonon energy (e.g.\
defect scattering, including isotopic mass defects) or transports
energy to the different parts of the Brillouin zone (e.g.\ umklapp
processes and boundary reflection/scattering). Both kinds of processes
constitute a deviation from the strictly harmonic regime (i.e.,
non-interacting phonons) and require either third-order contributions
to the interaction potential or separate modeling (e.g.\ Matthiessen’s
formula used for isotope scattering). The third-order interactions are
usually a leading contribution to phonon lifetimes. The next, fourth
order, the term is responsible for shifting the phonon frequencies 
since it mainly modifies the curvature of the energy surface. As
is obvious from Fig.~\ref{fig:phonons} and the SCPH calculations
reported above, the shifts due to the quartic (fourth-order)
components of the interaction potential are very small and limited to
the fine parts of the optical spectra near the X point.

\begin{figure}[!htbp]
\centering
\includegraphics[width=\figwidth]{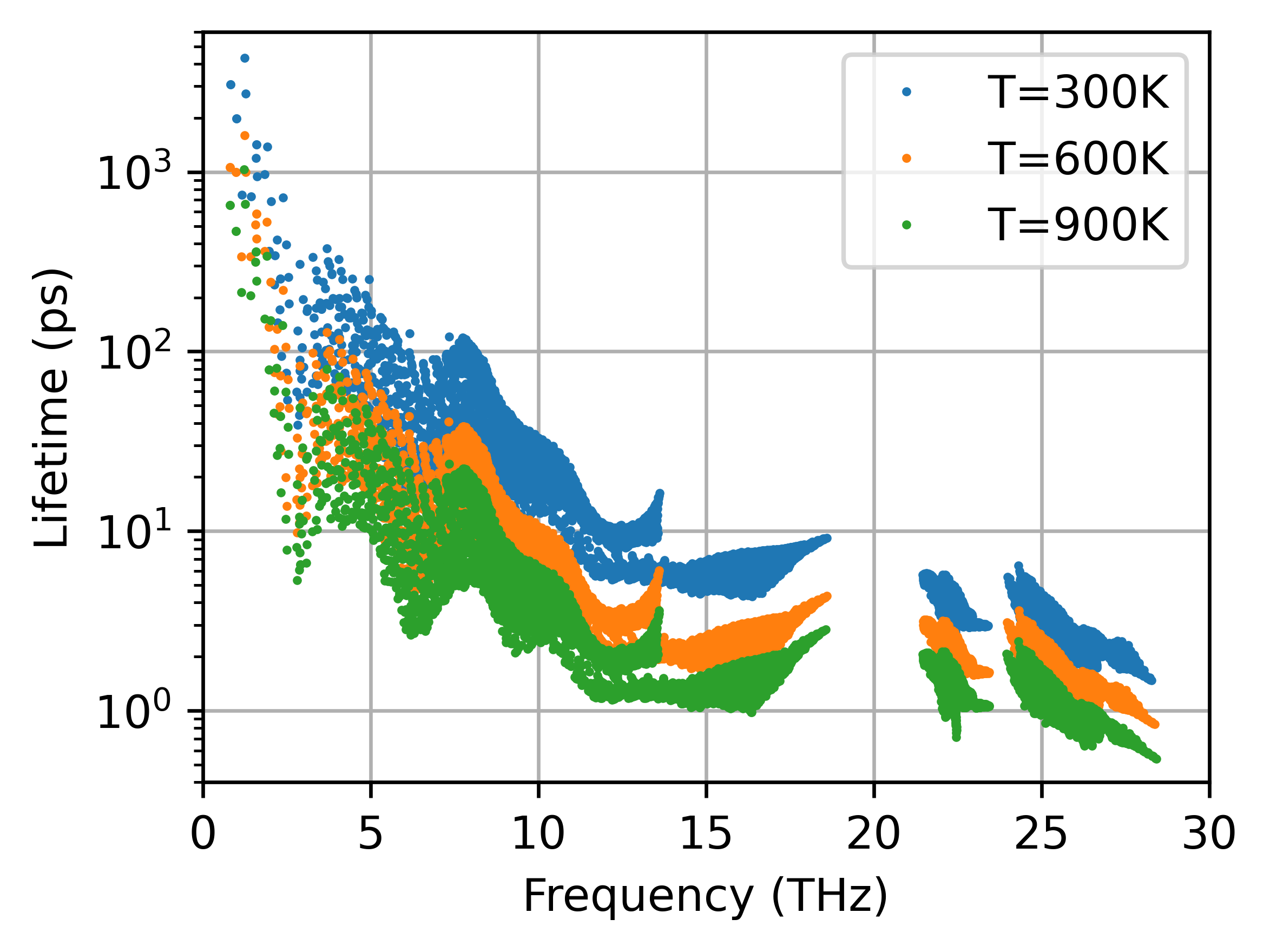}
\caption[]{Phonon lifetimes calculated with RTA at $T=300$, 600, 900 K, from the fourth-order potential model fitted to HECSS samples corresponding to the same temperature. Temperatures marked with colors (300~K -- blue, 600~K -- orange, and 900~K -- green).} 
\label{fig:tau-T}
\end{figure}

The influence of the third-order terms is more pronounced. The phonon
lifetimes calculated with Boltzmann transport equation and Relaxation
Time Approximation (BTE/RTA) plotted in Figure~\ref{fig:tau-T} as a
function of frequency for three different temperatures show a quite
strong temperature dependence and fairly short lifetimes (below 1~ps
for the top of the spectrum and high temperatures). This indicates
substantial third-order anharmonicity in the system.

\begin{figure}[!htbp]
\centering
\includegraphics[width=\figwidth]{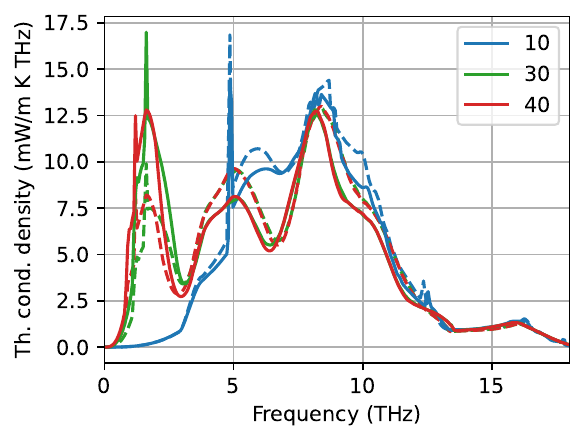}
\caption[]{Thermal conductivity spectral density for different k-grid sizes for the RTA procedure ($10^3$, $30^3$, $40^3$ respectively) and interaction model order (cubic - dashed line, quartic - solid line).}
\label{fig:thconspec}
\end{figure}

The phonon lifetime is not very easy to measure. This is just the first
step to determining the ultimate goal of our work, which is the thermal
conductivity coefficient for the material. The lattice thermal
conductivity \ref{eq:kappa} is a sum of frequency-dependent terms also
weighted by the Boltzmann factor. Thus, proper sampling of the
reciprocal space and energy scale is crucial for obtaining good accuracy
of the final result. The spectral decomposition of the thermal
conductivity contributions presented in Fig.~\ref{fig:thconspec} clearly
shows that selecting too coarse a grid for the calculation of the phonon
lifetimes removes low-frequency parts from the sum and may strongly
influence the final result. The data in Fig.~\ref{fig:thconspec}
indicates that the 30$\times$30$\times$30 points grid may already be
dense enough. For the final results, we have selected a slightly denser
grid of 40$\times$40$\times$40 points.
On the other hand, the comparison between two levels of the interaction
model (cubic -- dashed line and quartic -- solid line), indicates a much
smaller influence of this aspect. However, the quartic model seems to be
more stable between different grid spacings. Thus, the quartic model was
selected for the final calculation.

\begin{figure}[!htbp]
\centering
\includegraphics[width=\figwidth]{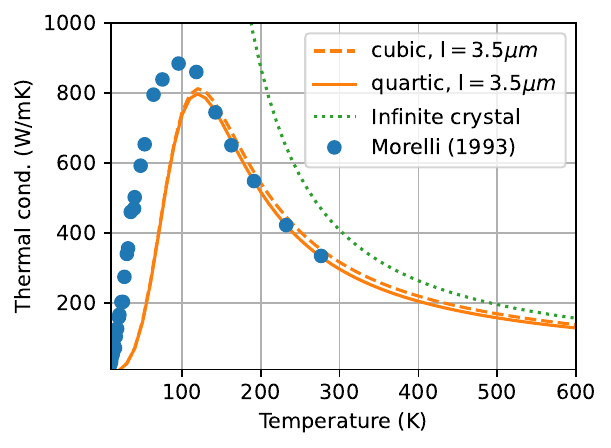}
\caption[]{Thermal conductivity coefficient calculated from samples
generated with temperatures $T=900$~K and k-grid size
40$\times$40$\times$40 in RTA calculation. The conductivity is
calculated for the 3.5$\mu$m grain size. The full line represents the
quartic interaction model, the dashed line cubic interaction model.
The dotted line represents the result for the infinite crystal. The
laboratory measurements (full dots) are from
\cite{morelli_carrier_1993}.}
\label{fig:thcond}
\end{figure}

The Peierls term \eqref{eq:kappa} contains contributions from all
wavelengths possible in the infinite crystal. This sum at low
temperatures is dominated by the contributions from the acoustic modes
with very low frequencies and high Boltzmann factor, and exhibits
infrared divergency common in such cases. This, obviously non-physical,
effect is shown in Figure~\ref{fig:thcond} by the dotted line.
The realistic calculation of the thermal conductivity of the material
requires including several finite-scale effects present in the real
crystal: isotope defect scattering, lattice defects scattering, and
finite crystallite size effects. The isotope scattering influence, in
the case of 3C-SiC, is vanishingly small (below $10^(-3)$).
The lattice defects play a much larger role, but their influence is
strongly sample-dependent and difficult to estimate. The final crystal
size, on the other hand, has a major influence on thermal conductivity
and is quite easy to include in the calculation --- by simply limiting
the wavelengths of the contributions to the size of the crystal.

The $\kappa(T)$ curve plotted in Fig~\ref{fig:thcond} was calculated
with the assumption of an average crystal size of $L=3.5
\mu m$. The comparison with experimental data from
\cite{morelli_carrier_1993} shows remarkably good agreement with the
measurements above 120~K. The low-temperature discrepancy can be
attributed to scattering on lattice imperfections --- as they influence
the curve similarly as the finite size of the crystal. Furthermore, a
strong influence of defects should be expected due to the known
difficulty of growing 3C-SiC monocrystals. Unfortunately, the
experimental data contains no information about the grain size or
defect concentration in the sample.

As we mentioned discussing Fig~\ref{fig:thconspec}, the degree of the
interaction model plays rather a minor role in this type of
calculation. This is confirmed by the insignificant difference between
the results from the cubic model (dashed line in
Fig~\ref{fig:thcond}) and the quartic one (solid line).

\section{Conclusions}\label{sec:conclusions}

The results presented above demonstrate the potential
of the proposed HECSS approach to generate a faithful representation
of the system in thermal equilibrium at elevated temperatures with only
a very small number of rejected DFT calculations -- thus very high
efficiency. The proposed new method of evaluation of sample weights
further improved the effectiveness of the original HECSS approach
\cite{jochym_high_2021} and added the ability to shift the target
temperature within the sampled energy range. The data from such
samplings allowed us to construct the high-accuracy
anharmonic (quartic) interaction model and calculate the
lattice thermal conductivity of 3C-SiC, which closely matches
experimental data above T=120~K. The extension of the results to
lower temperatures is limited only by further post-processing of data
from the interaction model to include lattice imperfection and
finite-size effects, which play an important role in low-temperature
thermal transport.

\begin{acknowledgments}
The authors would like to express their gratitude to Krzysztof Parlinski, Przemys{\l}aw Piekarz, Andrzej M. Ole{\'s}, and Ma{\l}gorzata Sternik for inspiring and fruitful discussions.
This work was partially supported by the National Science Centre (NCN, Poland) under grant UMO-2014/13/B/ST3/04393.
\end{acknowledgments}
\bibliography{refs.bib}
\end{document}